\documentclass[letterpaper, 10 pt, conference]{ieeeconf}  

\IEEEoverridecommandlockouts                              
\usepackage{graphics} 
\usepackage{epsfig} 

\usepackage{caption} 
\usepackage{subcaption}

\usepackage{amsmath} 
\usepackage{amssymb}  
\usepackage{mathtools}

\newtheorem{definition}{Definition}

\title{\LARGE \bf
Fast Physics-Informed Model Predictive Control Approximation for Lyapunov Stability
}

\author{Josue N. Rivera$^1$, Jianqi Ruan$^2$, XiaoLin Xu$^1$, Shuting Yang$^1$, Dengfeng Sun$^1$ and Neera Jain$^2$
\thanks{$^1$ School of Aeronautics and Astronautics, Purdue University, West Lafayette, Indiana IN, USA.
        {\tt\small \{river264, xu1103, yang1198, dsun\}@purdue.edu}.}%
\thanks{ $^2$ School of Mechanical Engineering, Purdue University, West Lafayette, Indiana IN, USA.
        {\tt\small \{ruan27, neerajain\}@purdue.edu}.}
}

\begin{document}

\maketitle
\thispagestyle{empty}
\pagestyle{empty}

\begin{abstract}
At the forefront of control techniques is Model Predictive Control (MPC). While MPCs are effective, their requisite to recompute an optimal control given a new state leads to sparse response to the system and may make their implementation infeasible in small systems with low computational resources. To address these limitations in stability control, this research presents a small deterministic Physics-Informed MPC Surrogate model (PI-MPCS). PI-MPCS was developed to approximate the control by an MPC while encouraging stability and robustness through the integration of the system dynamics and the formation of a Lyapunov stability profile. Empirical results are presented on the task of quadcopter landing. They demonstrate a rapid and precise MPC approximation on a non-linear system along with an estimated two times speed up on the computational requirements when compared against an MPC. PI-MPCS, in addition, displays a level of stable control for in- and out-of-distribution states as encouraged by the discrete dynamics residual and Lyapunov stability loss functions. PI-MPCS is meant to serve as a surrogate to MPC on situations in which the computational resources are limited. 

\end{abstract}

\section{INTRODUCTION}

Model Predictive Control (MPC) has emerged as a leading control technique due to its ability to consider both present and future implications of control actions. However, the computational demands of MPC can pose significant challenges, particularly in dynamic environments where frequent optimizations can strain systems with limited computational resources~\cite{holkar2010overview}. The iterative nature of MPC also introduces latency, necessitating sufficient time between control actions to accommodate the optimizations.

In light of these challenges, innovative solutions have been proposed to alleviate the computational demands of MPC across various systems. Physics-Informed Neural Network (PINN)-based MPC is one such promising avenue, previously explored in multi-link robot manipulators~\cite{NMPC}. This methodology synergies traditional MPC's strengths with data-driven techniques, crafting a hybrid that is both robust and adaptive~\cite{RAMP}. While various strategies like the Neural-Lander have emerged for specific use-cases like quadcopter descent~\cite{neural-lander}, the overarching challenge remains - ensuring computational efficiency without compromising on the depth of data and system dynamics.

In the extensive narrative of control problems using machine learning, significant strides have been made in embedding dynamics and physics knowledge into neural network-based learning processes to control a dynamical system. For instance, Markus Quade et al. elucidated the significance of explainable AI in the realm of machine learning control and spotlighted symbolic regression as a pivotal tool to infer the optimal control of a dynamical system~\cite{MLC_foundation}. Subsequent research by Shmalko and Diveev provided a deeper insight into control synthesis, leveraging and evaluating several symbolic regression techniques for obtaining the stabilization of a dynamical system~\cite{MLC_evolution}. A pivotal contribution in this realm was the introduction of advanced loss metrics for oscillatory data by Mathies Wedler et al., emphasizing refining loss functions tailored for systems with nonlinear dynamics~\cite{MLC_advancement}.

In response to these challenges and inspired by the idea of integrating dynamics and physics information into a neural network learning process, a strategy commonly attributed to Physics-Informed Neural Networks (PINNs), the research extends the thought to zero-order hold control strategies and presents a surrogate model for MPCs that approximates the stability control produced by the algorithm while encouraging a stable model through its objective function and at a fraction of an MPC’s computational need. The neural network serves as a complete surrogate model that observes a system state and returns the appropriate control input. These feats are achieved by including an approximated system’s dynamics and the novel creation and integration of a novel Lyapunov stability profile into the learning process. As a prominent behavior of modern quadcopters and as a well-studied stability problem, a case study on the automated stable decent of a quadcopter is presented, and our Physics-Informed MPC Surrogate (PI-MPCS) model performance when compared against an MPC.

\section{Preliminaries}

\subsection{Notation}

Let us define the discrete zero-order hold control time index $k$ at which a new state of a system $s$ is presented to a controller, and the resulting control response as $u_c$ and $\hat{u}_c$ by the MPC and PI-MPCS, respectively. Denote the time difference between new control signal responses as $\Delta T_c$. If the new control state is a result of the control generated by the MPC, let us define it as $s_+$. Alternatively, if the new state was produced as a consequence of the control signal by PI-MPCS, let us define it as $\hat{s}_+$. The state-space representation of such system is described by

\begin{equation}
    \label{eq:state-space}
    \dot{s} = f(s, u)
\end{equation}

where $u$ may be either zero-order hold $u_c$ or $\hat{u}_c$.

Let $D$ represent a list of tuples with $N_D$ elements formed from a collection of a quadcopter's trajectories when controlled by an MPC. Each tuple $(s, u_c, s_+)_i$ holds a state observed by the MPC $s$, the resulting control response $u_c$, and the next observed state $s_+$ after $\Delta T_c$. 

\subsection{MPC Reference Trajectory}

 A finite set of reference trajectories $D$ is collected that hold state and input information of the automated MPC system. These trajectories are generated from an initial state in a bounded set $\hat{\mathrm{H}} \subset \mathbb{R}^n$. All states within $\hat{\mathrm{H}}$ are denoted as initial in-distribution states supported by PI-MPCS. 

\begin{definition}[In-distribution states]
  A state $s$ is an in-distribution state if $s \in \mathrm{H} = Conv(\{s | s \in D\})$; the convex hull of all that states in the recorded trajectories $D$. The set of initial in-distribution states $\hat{\mathrm{H}} \subseteq \mathrm{H} \subset \mathbb{R}^n$.

\end{definition}

\begin{definition}[Out-of-distribution states]
  A state $s$ is an out-of-distribution state if $s \notin \mathrm{H}$.
\end{definition}

In a stability problem simulation, the MPC observes the states and provides a control input at a rate of $\Delta T_c$. At each instance of the observation and response, an element $(s, u_c, s_+)_i$ of $D$ is stored with $s$ and $s_+$ being sequential states resulting from $u_c$ being applied to the system \eqref{eq:state-space}.  

\subsection{Surrogate Models}

PI-MPCS and the surrogate models presented are described as a deterministic multilayer perceptron controller (i.e., a fully connected neural network) $h: \mathbb{R}^n \rightarrow \mathbb{R}^m$ that maps a system state $s$ to a control vector $\hat{u}_c$ as restricted on the MPC and described by

\begin{equation}
\label{eq:pi-mpcs}
\begin{split}
    \hat{u}_c = h(s; \mu) &= (h_k \circ \sigma \circ ... \circ h_2 \circ \sigma \circ h_1) (s) \\
    h_j(z) &=  W_j z + b_j
\end{split}
\end{equation}

where $\mu = \{W_1, b_1, ..., W_k, b_k\}$ is the set of all the learnable parameters of the model through optimization and $\sigma(\cdot)$ is a non-linear activation function.


\section{Physics-Informed MPC Surrogate}

\subsection{Control Mimicry}

As PI-MPCS and other surrogates (\ref{eq:pi-mpcs}) yields a control response $\hat{u}_c$ and the expected control response $u_c$ by an MPC is known for each state $s \in D$, an organic and the first loss term of the objective function is the \textit{control loss} function (\ref{eq:l1}).

\begin{equation}
\label{eq:l1}
L_{1} = \smashoperator{\sum_{s(i), u_c(i) \in D}} \frac{[u_c(i) - h(s(i); \mu)]^2}{N_D}
\end{equation}

where $s(i)$ and $u(i)$ refers to the state $s$ and control $u$ of the $i^{th}$ element of $D$, respectively, and $h(\cdot)$ is PI-MPCS as defined in (\ref{eq:pi-mpcs}). A neural network surrogate model optimized for just this loss function will be used as a benchmark model.

\subsection{Approximated Dynamics}

To integrate information about the system dynamics with a computationally efficient approach, the Euler method is employed to approximate the future state $\hat{s}_+$ given the current state of the system $s$ and the control by PI-MPCS $\hat{u}$.
\begin{equation}
\label{eq:s+}
    \hat{s}_+(s, \hat{u}) = s + \Delta T_c \cdot f(s, \hat{u})
\end{equation}

where $s(i)$ and $s_+(i)$ refers to the state $s$ and future state $s_+$ of the $i^{th}$ element of $D$, respectively. $f$ is a first-order state space representation of the dynamics defined in (\ref{eq:state-space}). Note that $\hat{s}_+(s, \hat{u})$ may just be referred by the notation $\hat{s}_+$ moving forward. The approximation of the future states $\hat{s}_+$ naturally leads to the \textit{discrete dynamics residual loss} function (\ref{eq:l2}) for the controller optimization.

\begin{equation}
\label{eq:l2}
L_{2} = \smashoperator{\sum_{s(i), s_+(i) \in D}} \frac{[s_+(i) - \hat{s}_+(s(i), h(s(i); \mu))]^2}{N_D}
\end{equation}

where $\hat{s}_+(\cdot)$ is the future state approximation described in (\ref{eq:s+}). The loss function intends to adjust the controls $\hat{u}_c$ for a future state $\hat{s}_+$ that minimizes the mean squared error (MSE) with the future state observed by the MPC $s_+$ when given the same state $s$.

\subsection{Lyapunov Stability Profile}

When observing a system controlled by an MPC or PI-MPCS for stability, such system can be classified as an autonomous system. By Lyapunov's second method for stability, an autonomous system is stable in the sense of Lyapunov at $s$ if in a neighborhood $W$ of $s$ there exists a Lyapunov function $V(s):W \rightarrow R^+$, and the $\Delta(V(s)) = V(s_+) - V(s) < 0$ for $s_+ \in W$ \cite{roberto2017}. 

Consider the quadratic Lyapunov function,

\begin{equation}
    \label{eq:V(z)}
    V(s) = s^T P s 
\end{equation}

where $P$ is positive definite. Given the states, $s$ and future state $s_+$ of $D$, an approximation of $V(s)$ for a system controlled by an MPC can be obtained by finding a solution P that holds to the Lyapunov stability criterion. The following optimization (\ref{eq:p-optimization}) may be performed to obtain an estimation of $V(s)$ through finite samples

\begin{equation}    
\begin{split}  
    \label{eq:p-optimization}
    \min_{P \in S^{n \times n}} \; &f(P) = \smashoperator{\sum_{s(i), s_+(i) \in D}} \max(V(s_+(i)) - V(s(i)), 0) \\
    \textit{s.t.} \; &P \succ 0
\end{split}
\end{equation}

where $s(i)$ and $s_+(i)$ refers to the $i^{th}$ state $s$ and future state $s_+$ of $D$, and $S^{nxn}$ denotes the set of $n \times n$ real symmetric matrices. The resulting solution $P$ of the optimization can be defined as the stability profile.

\begin{definition}[Lyapunov stability profile]
  A solution to the optimization (\ref{eq:p-optimization}) $P$ is defined as a Lyapunov stability profile given a collection of trajectories $D$.
\end{definition}

Given a profile $P$, the \textit{Lyapunov stability loss} function can be formed that encourages similar stability properties for the autonomous system when guided by PI-MPCS

\begin{equation}
    L_{3} = \smashoperator{\sum_{s(i) \in D}} \frac{[\max\left(V(\hat{s}_+(i)) - V(s(i)),0\right)]^2}{N_D}
\end{equation}

where $\hat{s}_+(i)=\hat{s}_+(s(i), h(s(i); \mu))$ is the future state approximation described in (\ref{eq:s+}) with input (\ref{eq:pi-mpcs}), and $V(\cdot)$ is the quadratic Lyapunov function (\ref{eq:V(z)}) given the profile $P$ from solving (\ref{eq:p-optimization}).

As the density of the states $s\in D$ in the in-distribution set $\mathrm{H}$ may be clustered, and the variance of the states near the origin may be small (a critical neighborhood for the stability problem), auxiliary states are generated to augment the diversity of the states. Given the region $\Tilde{\mathrm{H}} \subset \mathrm{H}$ where the density of the states in $D$ are low and the neighborhood of feasible states near the origin $\Tilde{H}_0$, random states are sampled to form a finite set of auxiliary states $\Tilde{\mathrm{S}} \subset \{\Tilde{\mathrm{H}} \cup \Tilde{H}_0\}$. In particular, states in $\mathrm{H}$ and $\Tilde{H}_0$ are sampled uniformly and combined to form the auxiliary states $\Tilde{\mathrm{S}}$.

The introduction of these auxiliary states results in a new augmented \textit{Lyapunov stability loss}

\begin{equation}
    \Tilde{L}_{3} = \smashoperator{\sum_{s(i) \in \{D_s \cup \Tilde{S}\}}} \frac{[\max\left(V(\hat{s}_+(i)) - V(s(i)),0\right)]^2}{N_{D_s} + N_{\Tilde{S}}}
\end{equation}

where $N_{\Tilde{S}}$ is the number of points in the set $\Tilde{S}$ and $s(i) \in \{D_s \cup \Tilde{S}\}$ refers to each state in the union of the states $s \in D$ and the auxiliary states $\Tilde{S}$. Note that only the \textit{Lyapunov stability loss}, so far, may be augmented as it is the only loss function that entirely relies on state-only information for its evaluation. Note that these new auxiliary states are not used in the optimization for the for the Lyapunov stability profile \eqref{eq:p-optimization}.  

A final \textit{feasibility loss} function may be introduced, but necessary, that takes advantage of the auxiliary states $\Tilde{S}$ and supports the avoidance of unfeasible states

\begin{equation}
    \Tilde{L}_{4} = \smashoperator{\sum_{s(i) \in \{D_s \cup \Tilde{S}\}}} \frac{g(s(i))}{N_{D_s} + N_{\Tilde{S}}}
\end{equation}

where is $g(s(i))$ is a problem dependent penalty function. As unfeasible states may exist in the trajectory dataset $D$, the \textit{feasibility loss} discourages such states.

\subsection{Convexity of Lyapunov Stability Profile Optimization}

Lyapunov stability profile optimization \eqref{eq:p-optimization} with respect to $P$ can be proven to always have a solution given a set of trajectories $D$ as the optimization is convex.

\noindent\textit{Proof.}


For the specific Lyapunov function in \eqref{eq:V(z)}, \eqref{eq:p-optimization} can be written as 
\small
\begin{align}
    f(P) &= \sum^I_{i=1} \max(s_+(i)^T P s_+(i) - s(i)^T P s(i),0)
\end{align}
\normalsize
Further split $f(P)$ into two parts, $f_a(P)$ and $f_b(P)$, expressed in \eqref{eq-f1} and \eqref{eq-f2}.
\small
\begin{align}
    f(P) &= f_a(P) + f_b(P)\\ \label{eq-f1}
         &= \sum_{a\in I} \max(s_+(a)^T P s_+(a) - s(a)^T P s(a),0) \\ \label{eq-f2}
         &+ \sum_{b\in I} \max(s_+(b)^T P s_+(b) - s(b)^T P s(b),0) 
\end{align}
\normalsize
where $a\in I_a \subseteq I ; s_+(a)^T P s_+(a) - s(a)^T P s(a) \leq 0$, and $b\in I_b \subseteq I ; s_+(b)^T P s_+(b) - s(b)^T P s(b) > 0$.

Since $s_+(a)^T P s_+(a) - s(a)^T P s(a) \leq 0 $, $f_1(P) = 0$. $f_a(P) = 0$ is both convex and concave because the line segment between any two points on the graph of $f_a=0$ lies on the graph \cite{proof,proof2}. 

Since $s_+(b)^T P s_+(b) - s(b)^T P s(b) > 0$, the part in \eqref{eq-f2} can be rewritten as $f_b(P) = \sum_{b\in I} s_+(b)^T P s_+(b) - s(b)^T P s(b)$. In this case, the optimization problem can be formulated as a Semidefinite Programming (SDP) problem. 

Define $B \leq I$ is the number of samples in $I_b$. Then define $X = [s(b_1),...,s(b_B)]$, and $X_+ = [s_+(b_1),...,s_+(b_B)]$

\small
\begin{align}
    f_2(P) &= \sum_{b\in I} (s_+(b)^T P s_+(b) - s(b)^T P s(b)) \\ \nonumber
           &= Tr(X_{+}^T P X_{+} - X^T P X) \\ \nonumber
           &= Tr(X_{+}^T P X_{+}) - Tr(X^T P X) \\ \nonumber
           &= Tr(X_{+} X_{+}^T P) - Tr(X X^T P) \\ \nonumber
           &= Tr(X_{+} X_{+}^T P - X X^T P) \\ \nonumber
           &= Tr((X_{+} X_{+}^T - X X^T) P) \\ \nonumber
           &= Tr(CP)
\end{align}
\normalsize
where $C = X_{+} X_{+}^T - X X^T $. $C$ is a $n \times n$ real symmetric matrix.
The P-optimization problem can be reformulated as below
\begin{align}
    \min_{P \in S^{n \times n}} \; & Tr(CP)\\
    \textit{s.t.} \; &P \succ 0
\end{align}
The reformulated problem matches the definition of a standard SDP problem \cite{sdp} but with a stricter constraint that the decision variable, P matrix, needs to be positive definite instead of positive semidefinite. As proved by \cite{sdp,sdp2}, an optimization problem formulated as SDP is convex. By now, $f_1(P)$ and $f_2(P)$ are proved to be convex. The convexity of the P-optimization problem is proved.

\subsection{Objective Function and Optimization}

Given the various loss functions, a net objective function can be formed that encapsulates desired behaviors for the surrogate model

\begin{equation} 
    \label{eq:objc}
    L = w_1 L_1 + w_2 L_2 + w_3 \Tilde{L}_3 + w_4 \Tilde{L}_4
\end{equation}

where $w_1, w_2, w_3$ and $w_4$ serve as adjustable hyperparameters to scale and penalize some loss terms more than others. 

With the definition of the objective function, one can perform the following optimization to obtain the parameters of PI-MPCS

\begin{equation}    
\begin{split}  
    \label{eq:pi-mpcs-optimization}
    \min_{\mu} \; & L (h(\cdot; \mu), D, P) 
\end{split}
\end{equation}

where $L(\cdot)$ is the evaluation of the objective function (\ref{eq:objc}) with respect to PI-MPCS $h(\cdot)$ (\ref{eq:pi-mpcs}) and its parameters $\mu$, the collection of trajectories $D$ generated from the simulations of the system controlled by the MPC, and the Lyapunov stability profile $P$ determined by (\ref{eq:p-optimization}). 

PI-MPCS refers to physics-informed surrogate model \eqref{eq:pi-mpcs} trained with at least terms $L_1$, $L_2$ and $L_3$ in the objective function during optimization \eqref{eq:pi-mpcs-optimization}, while a physics-informed model is any surrogate model trained with $L_2$ or $L_3$.

\section{RESULTS}

\subsection{Case Study}

To evaluate the performance of PI-MPCS in an stability problem, a case study on quadcopter landing is presented. The quadcopter is modeled as a rigid body. From the Newton–Euler equations, one can describe the system as

\begin{equation}
\label{eq:system-ode}
\begin{split}
    \ddot{x} &= -\sin(\theta)(u_R + u_L)/m \\
    \ddot{y} &= (\cos(\theta)(u_R + u_L)-g)/m \\
    \ddot{\theta} &= L(u_R - u_L)/J
\end{split}
\end{equation}

where $m$, $L$, $J$ and $g$ are the mass, half-length, inertia, and gravity in the system, $u_R$ and $u_L$ are the right and left thrust force, and $[u_R, u_L]^T = u = u_e + u_s + u_c$ is the net control signal applied to the system by an MPC. If PI-MPCS applies the control, then the net control is expressed as $\hat{u} = u_e + u_s + \hat{u}_c$. The controls $u_e$ and $u_s$ are described by

\begin{equation}
    u_e = [mg/2, mg/2]^T
\end{equation}
and
\begin{equation}
    u_s = \kappa s
\end{equation}

where $s = [x,y,\theta,\dot{x},\dot{y},\dot{\theta}]^T$ is the state vector. $u_e + u_s$ is a linear state feedback stabilizer with static gain $\kappa$. The stabilizer $u_s$ drives the tilt angle $\theta$ and all first-order time derivative states to zero and keeps the drone in a hovering position when no other control signals are present. From (\ref{eq:system-ode}), a state space representation of this system can be described as

\begin{equation}
    \label{eq:state-space-quad}
    \dot{s} = f(s, u) = 
    \begin{bmatrix}
        s_4 \\
        s_5 \\
        s_6 \\
        -\sin(s_3)(u_1 + u_2)/m \\
        (\cos(s_3)(u_1 + u_2)-g)/m \\
        L(u_1 - u_2)/J
    \end{bmatrix}
\end{equation}

The task is to construct a surrogate PI-MPCS controller that replicate the ``human level'' control $u_c$ produced by an MPC. PI-MPCS must provide a similar control as the MPC that drives the drone from a location inside a initial position region (in-distribution initial states).

In the case study of the quadcopter landing problem, the \textit{feasibility loss} can be defined as

\begin{equation}
    \Tilde{L}_{4} = \smashoperator{\sum_{s(i) \in \{D_s \cup \Tilde{S}\}}} \frac{\max\left(\hat{y}_{+}(i),0\right)^2}{N_{D_s} + N_{\Tilde{S}}}
\end{equation}

where $\hat{y}_{+} (i)$ is the horizontal displacement of the approximated future state $\hat{s}_+(i) = \hat{s}_+(s(i), h(s(i); \mu))$ given the state $s$ of the $i^{th}$ element of $D_s \cup \Tilde{S}$. The loss function discourages negative horizontal displacement as it is unfeasible in the quadcopter landing problem. The formulation of the \textit{feasibility loss} relies on the understanding of the problem setting and it is not necessary for PI-MPCS.

\subsection{Implementation}

The reference trajectories $D$ collected for the test cases in the result section contain the observations and control inputs of the MPC sampled at a rate of 20Hz. Since there is no noise or disturbance considered, trajectories starting from the same initial condition are deterministic. $D$ contains 273 trajectories with each lasting 15 seconds and producing 300 samples. Each trajectory is given an initial condition $s_0$ in the in-distribution initial states set $\hat{\mathrm{H}} = \{ ( x,y,0,0,0,0) \in \mathbb{R}^2\times[0, 2\pi]\times \mathbb{R}^3: x \in [-2.5, 2.5], y \in [3.5, 6.5]\}$. The convex hull of all the trajectory states $s$ resulting from these initial states $s_0$ forms the \textit{in-distribution states} set $\mathrm{H}$.

PI-MPCS and the surrogates (\ref{eq:pi-mpcs}) consists of $k=4$ layers and the sigmoid activation functions $\sigma (z) = 1/(1+e^{-x})$. It contains a total of 312 learnable parameters optimized across 200 epochs iterations and initialized using Kaiming uniform distribution \cite{he2015delving}. The optimization algorithm Adam is used with the hyperparameters suggested by the authors and a learning rate of 1e-3 \cite{kingma2014adam}. The weights for the net objective function were set to $[w_1, w_2, w_3, w_4] = [1, 1000, 1, 100]$ in order to scale $L_1$ through $L_4$ to similar magnitudes.

In the simulation environment, PI-MPCS replaces the MPC controller, and the new states are presented at the same rate that was presented to the MPC $\Delta T_c = 0.05$ seconds. The rest of the simulation environment remains the same.

\subsection{In-Distribution Performance}

In order to demonstrate the effectiveness of each loss function and the auxiliary set, eight surrogate models are trained with the reference trajectories $D$. The controller solely trained with the \textit{control loss} $L_1$ is considered as the benchmark. The other 7 controllers are trained with different combinations of loss functions and with/without an auxiliary set. Four performance metrics are defined for performance comparison:
\begin{enumerate}
    \item \emph{Successful landing} - Considering the size of the simulated drone, if the controller drives the drone to the region $(x,y)\in[-0.5,0.5]\times[0,0.5]$ within 15 seconds and stays in this region to the end of simulation, then it is considered as a successful landing.
    \item \emph{Safe landing} - The landing simulation scenario requires the drone to not land underground. Thus, a landing trajectory of 15-second simulation, which does not have samples with negative $y$ components, is considered as a safe landing.
    \item \emph{Landing time} - The landing time metric is applicable to trajectories that are considered as successful landing. The landing time is defined as the first time step when the drone is in the defined landing region $(x,y)\in[-0.5,0.5]\times[0,0.5]$.
    \item \emph{Tracking error} - The tracking error metric is applicable to trajectories that are considered as successful landing. It is defined as a surrogate model's positional error in reference to that of the MPC for the same initial condition \eqref{eq-tracking_err}.
    \begin{equation} 
    \label{eq-tracking_err}
    \epsilon(k) = \sqrt{(x(k) - \hat{x}(k))^2 + (y(k) - \hat{y}(k))^2} 
\end{equation} 
\end{enumerate}

All surrogate model are evaluated by simulating 100 landing trajectories from random initial locations. The successful landing rate, the safe landing rate, the average landing time, and the average tracking error are presented in Tab.~\ref{tab-perf_comp}. Six out of Seven PI-MPCS have higher successful landing rates than the benchmark controller (No. 1.) In addition, five out of six PI-MPCS that are trained with the \textit{Lyapunov stability loss} $L_3$ show more improvements in the successful landing rate than Controller No. 2, which does not use $L_3$ for training. Effects of the \textit{feasibility loss} $L_4$ are shown by comparing Controller No. 3 against No. 4 and No. 6 against No. 7. The difference in each pair is whether the \textit{feasibility loss} $L_4$ is used in training. Controllers No. 3 and No. 6 trained with $L_4$ have higher safe landing rates than No. 4 and No.7, respectively. The PI-MPCS trained with an auxiliary set (No. 6-8) have higher successful landing rates, safe landing rates, and lower tracking errors, compared to their corresponding PI-MPCS trained without an auxiliary set (No. 3-5). However, the average landing times of No.6-8 are slightly longer than their corresponding PI-MPCS in No. 3-5.

\begin{figure}
    \centering
    \begin{subfigure}[b]{0.23\textwidth}
        \includegraphics[width=\textwidth]{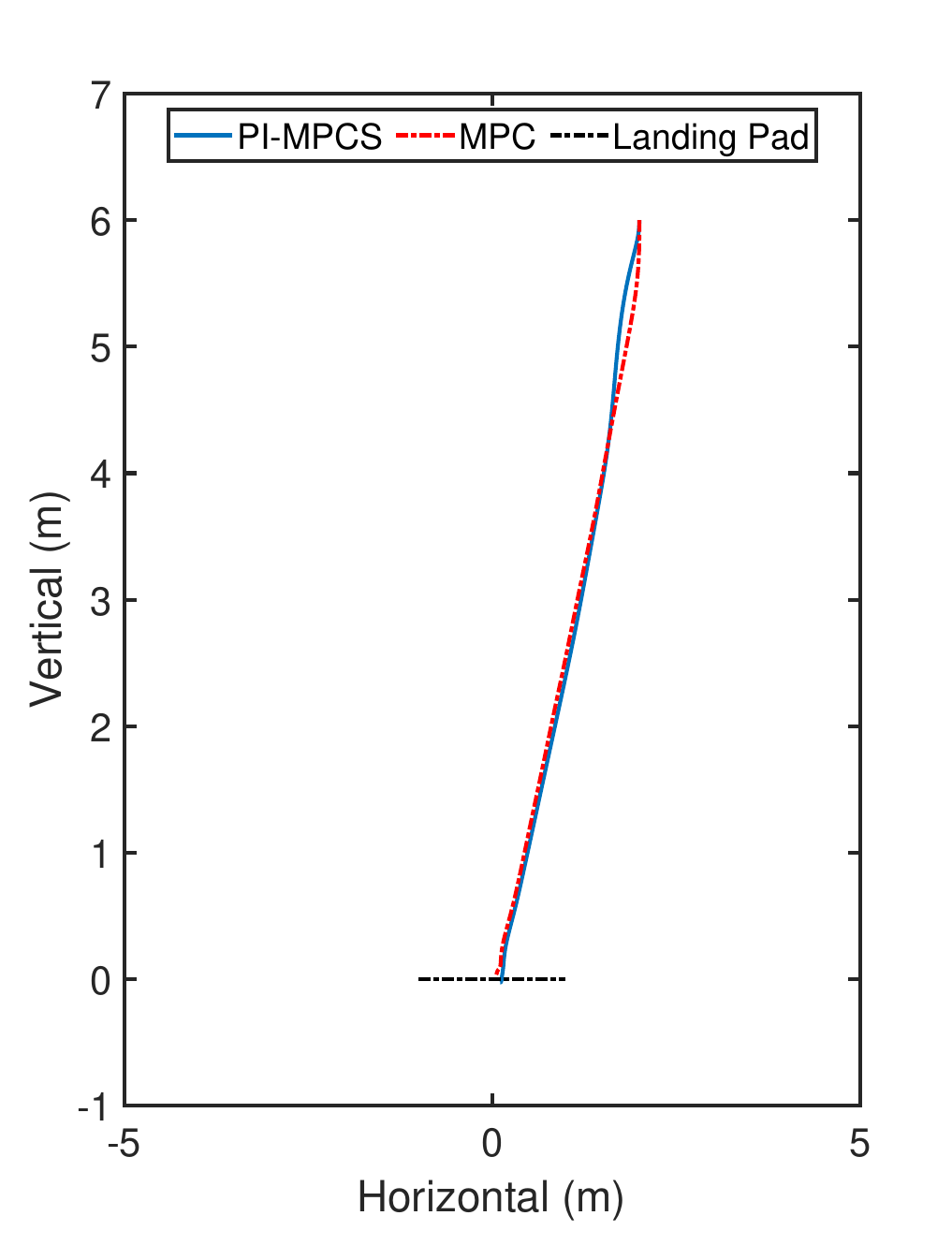}
        \caption[]%
        {{\small}}    
    \end{subfigure}
    \begin{subfigure}[b]{0.23\textwidth}  
        \includegraphics[width=\textwidth]{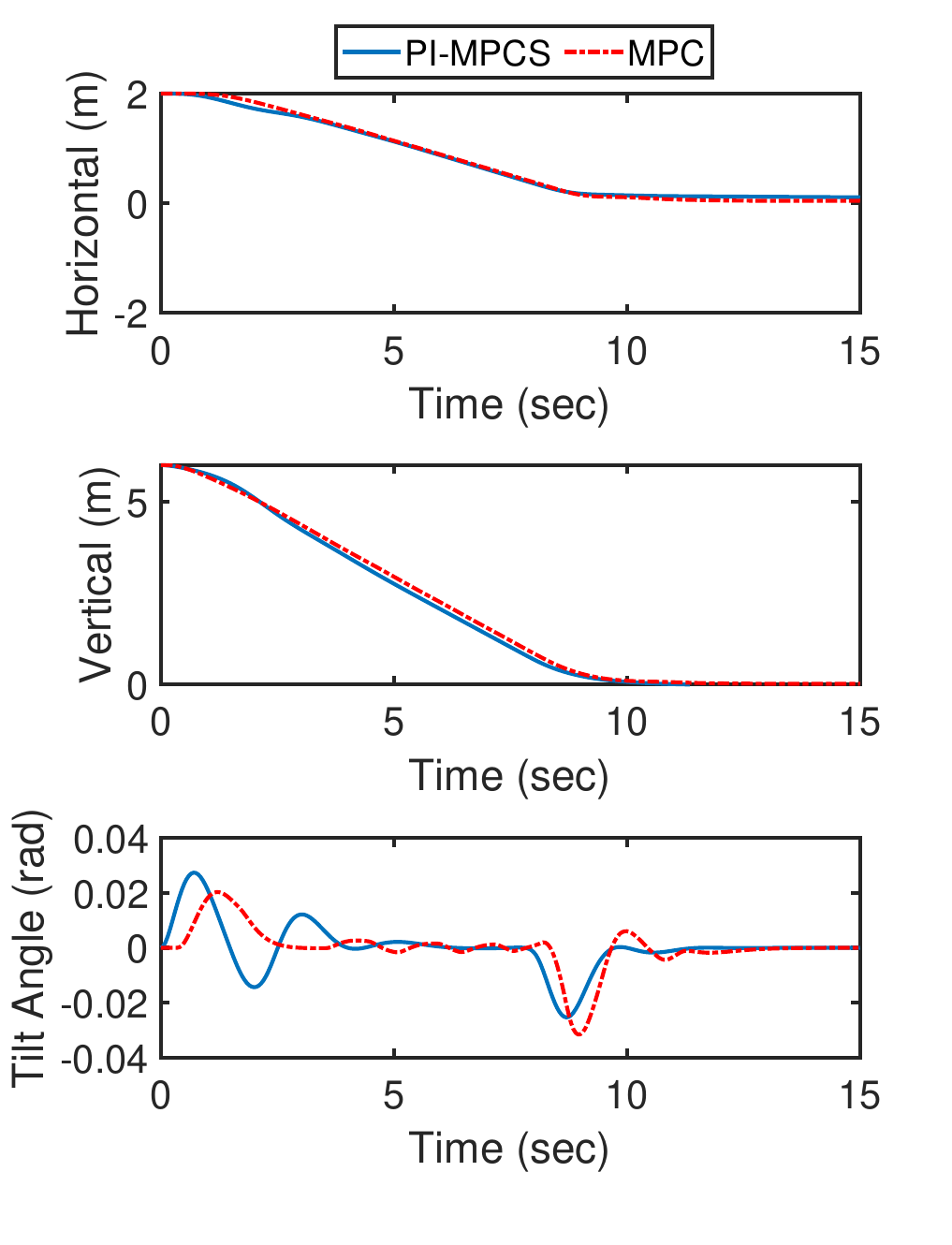}
        \caption[]%
        {{\small}}
    \end{subfigure}
    \caption[]
    {\small Test case 1 with initial state $s_0 = [2,6,0,0,0,0]$}
    \label{fig-track_case1}
\end{figure}

\begin{figure}
    \centering
    \begin{subfigure}[b]{0.23\textwidth}
        \includegraphics[width=\textwidth]{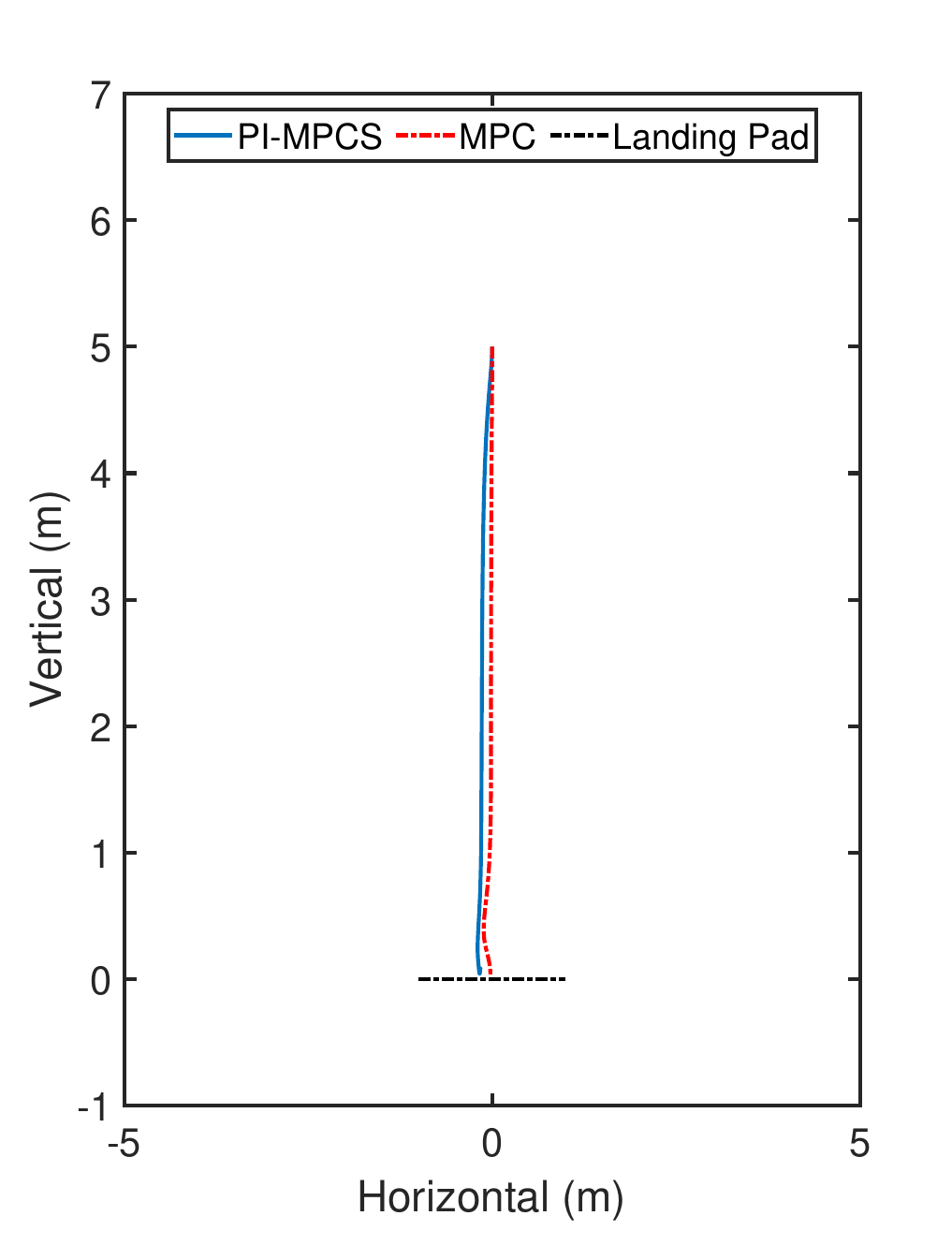}
        \caption[]%
        {{\small}}    
    \end{subfigure}
    \begin{subfigure}[b]{0.23\textwidth}  
        \includegraphics[width=\textwidth]{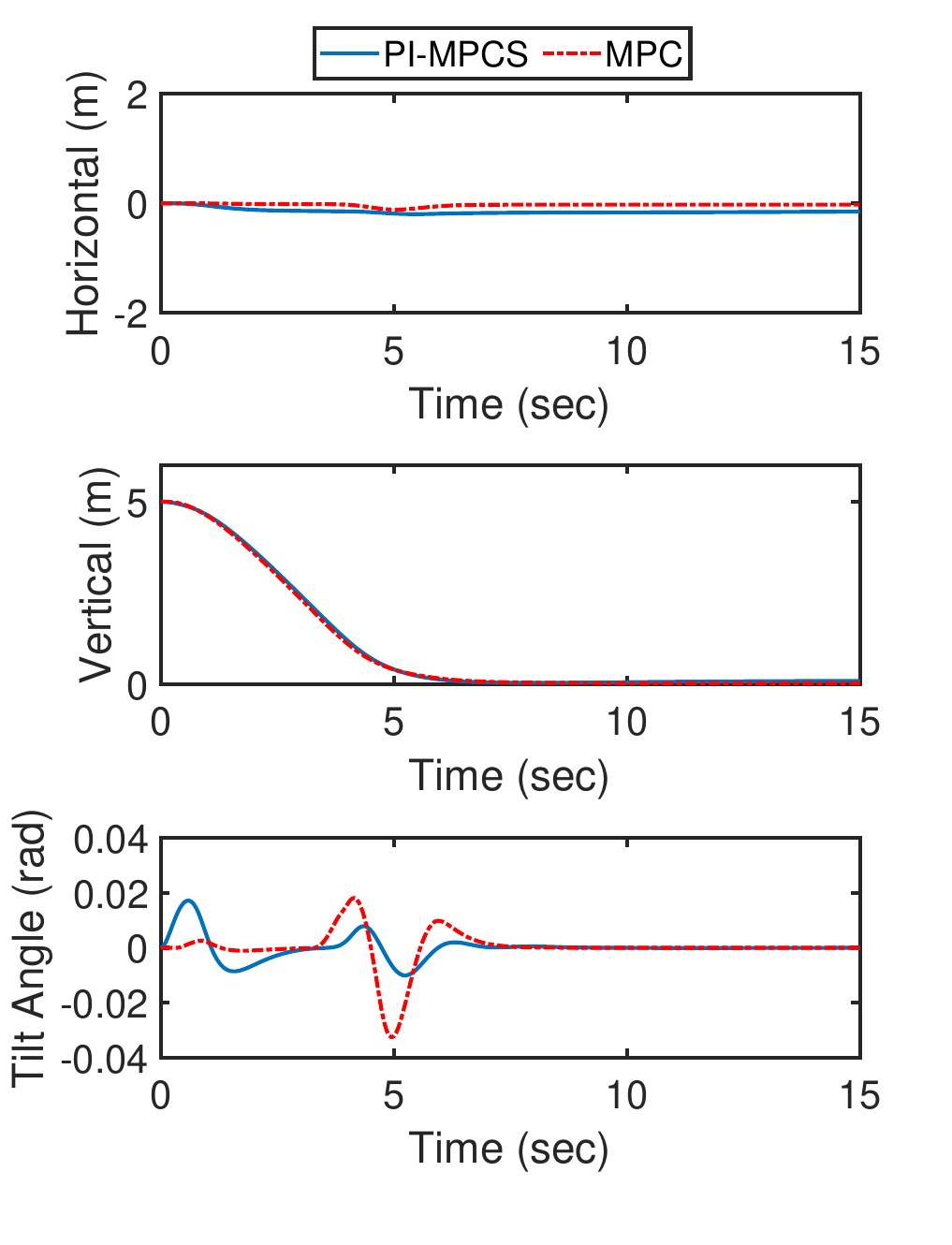}
        \caption[]%
        {{\small}}
    \end{subfigure}
    \caption[]
    {\small Test case 2 with initial state $s_0 = [0,5,0,0,0,0]$}
    \label{fig-track_case2}
\end{figure}

The best performance with respect to each metric are labeled in bold. Controller No. 4 has the highest successful landing rate and shortest average landing time. However, its safe landing rate is the lowest among all surrogate models. Comparatively, Controller No. 7 has the same successful landing rate and similar landing time and tracking error, but $19\%$ higher safe landing rate. Figures~\ref{fig-track_case1} and~\ref{fig-track_case2} show two landing trajectory comparisons between Controller No. 7 (PI-MPCS) and the MPC. The two cases show that no matter whether the initial location is strictly above the landing pad or not, the trajectory driven by PI-MPCS is similar to that driven by the MPC. The most noticeable difference is in the tilt angle; the PI-MPCS tends to have more aggressive tilt angle at the beginning, and the MPC tends to tune the tilt angle more aggressively when the drone is approaching the landing pad (i.e., $y = 0$).

There is also a limitation of this study as shown in Tab.~\ref{tab-perf_comp}. Although it is demonstrated that including the \textit{feasibility loss} $L_4$ can increase the safe landing rate of the PI-MPCS, none of the PI-MPCS has a safe landing rate higher than the benchmark (Controller No. 1.)  A potential explanation is that the reference trajectories $D$ have unfeasible samples, i.e., samples with negative $y$ component in their next-step state $s_+(i)$. For controllers trained with the \textit{discrete dynamics residual loss} $L_2$, the controllers learn the control inputs that drives the drone into the unfeasible state. For controllers trained with the \textit{Lyapunov loss} $L_3$, since the matrix $P$ is determined by the next-step state $s_+(i)$, the controllers also learn the control inputs that drives the drone into the unfeasible state to satisfy the Lyapunov stability criterion with respect to the certain $P$ matrix.

\begin{table}[ht]
\centering
\caption{The successful landing rate, the safe landing rate, the average landing time, and the average tracking errors of controllers. The best performance with respect to each metric is bold. The controllers marked (*) are our PI-MPCS.}
\label{tab-perf_comp}
\begin{tabular}{l|ll|cccc}
No. & Loss function     & \begin{tabular}[c]{@{}l@{}}Aux.\\ Set\end{tabular} & \multicolumn{1}{l}{Suc.} & \multicolumn{1}{l}{Safe} & \multicolumn{1}{l}{Time} & \multicolumn{1}{l}{Tracking} \\ \hline
1   & $L_1$             &                                                    & 0.76                     & \textbf{0.9}             & 6.65                     & 0.035                        \\
2   & $L_1,L_2$         &                                                    & 0.83                     & 0.84                     & 6.55                     & 0.051                        \\
3*   & $L_1,L_2,L_3,L_4$ &                                                    & 0.94                     & 0.72                     & 7.56                     & 0.455                        \\
4*   & $L_1,L_2,L_3$     &                                                    & \textbf{1}               & 0.42                     & \textbf{6.25}            & 0.015                        \\
5   & $L_1,L_3$         &                                                    & 0.73                     & 0.44                     & 6.47                     & 0.163                        \\
6*   & $L_1,L_2,\Tilde{L}_3,\Tilde{L}_4$  & Yes                                                & 0.96                     & 0.76                     & 7.58                     & 0.206                        \\
7*   & $L_1,L_2,\Tilde{L}_3$    & Yes                                                & \textbf{1}               & 0.61                     & 6.26                     & \textbf{0.013}               \\
8   & $L_1,\Tilde{L}_3$         & Yes                                                & 0.87                     & 0.53                     & 6.74                     & 0.043                       
\end{tabular}
\end{table}

\begin{figure}
  \centering
  \includegraphics[width=1\linewidth]{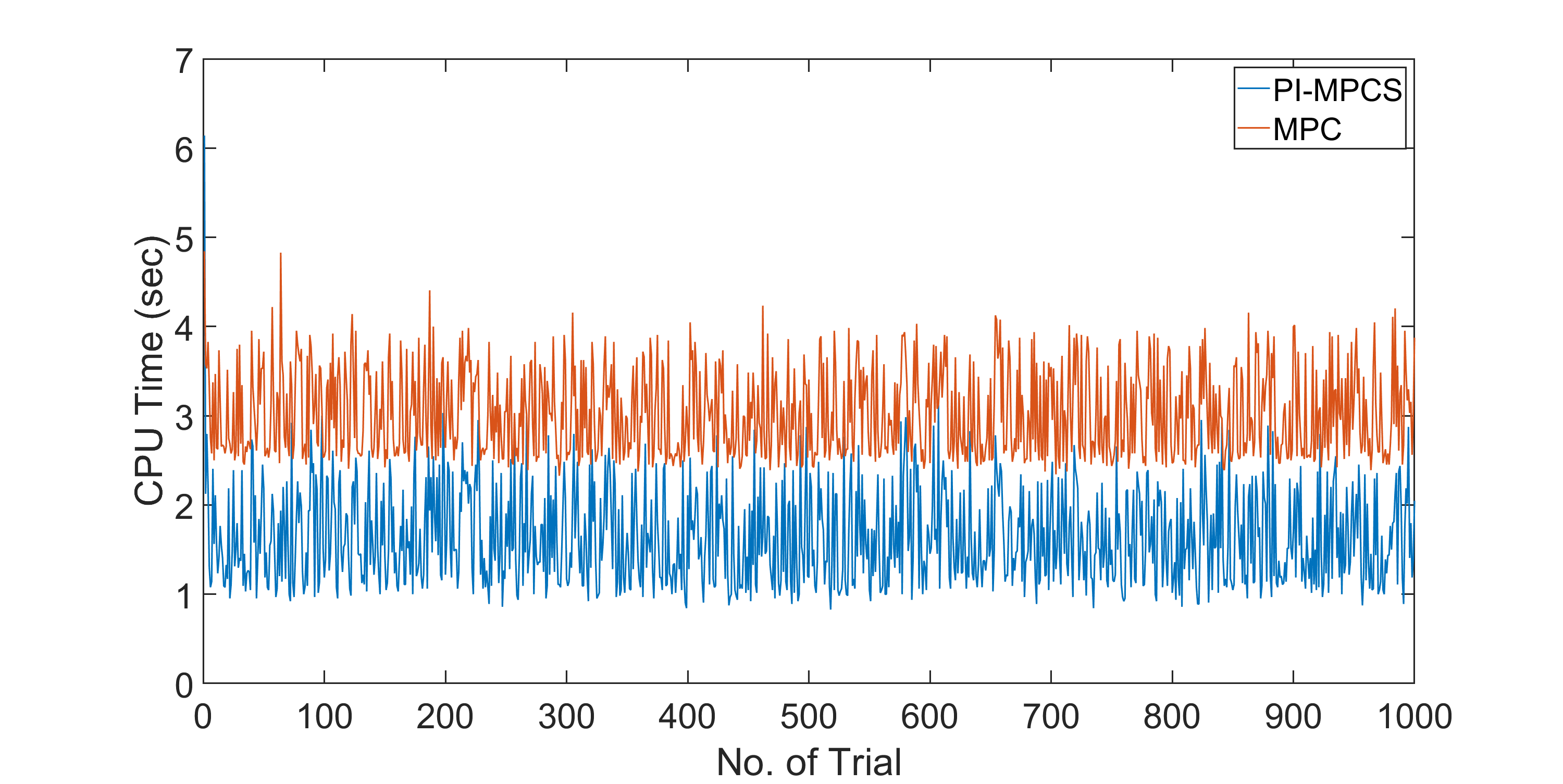}
  \caption{CPU time comparison between MPC (mean 3.2037 secs, standard deviation 0.5653 sec) and PI-MPCS (mean 1.7344 sec, standard deviation 0.5370 sec)}
  \label{fig-time_comp}
\end{figure}

\subsection{Computational Time Performance}

To showcase the model performance across a variety of initial states, the Monte Carlo method was used to find an estimation of the error distribution. The execution time consumed by the CPU was also recorded. Results are presented in Fig. \ref{fig-time_comp}. By replacing the controller from an MPC to PI-MPCS, it is observed that the expected CPU time consumption is reduced by $45\%$. In these simulations, the controller was the only component that changed, and the rest of the components remained the same.




\section{CONCLUSION}

This paper presents PI-MPCS -- a Physics-Informed Model Predictive Control Surrogate model that approximates an MPC's control policy. Reference trajectories generated by the MPC are used to train PI-MPCS for approximation and similar stability profile. The model is trained using: a \textit{control loss function} that emphasizes control input similarity between the MPC and PI-MPCS, a \textit{discrete dynamics loss} that motivates similarities in the consequent state, a novel \textit{Lyapunov stability loss} that encourages a similar stability profile, and an optional \textit{feasibility loss} that discourages unfeasible states. In addition, an auxiliary set is introduced to further enforce the PI-MPCS to follow the stability and the feasibility criteria. A case study on stable quadcopter descent is presented to demonstrate the effectiveness of each physics-informed loss function as well as the auxiliary set. The main empirical finding were:

\begin{enumerate}
    \item Controllers trained with physics-informed loss functions have higher success rates in stable decent than standard neural network surrogate which is only trained with the \textit{control loss}.
    \item Controllers trained with the \textit{Lyapunov loss} have higher success rates in stable decent than those trained without using the \textit{Lyapunov loss}.
    \item When using the \textit{feasibility loss} or not is the only variable, controllers trained with the \textit{feasibility loss} have higher safe landing rates.
    \item When using the auxiliary set or not is the only variable, controllers trained with the auxiliary set have higher safe landing rates and higher success rates in stable decent while achieving smaller tracking error.
\end{enumerate}

 Greater computationally efficient when compared against an MPC is also observed. PI-MPCS is a compact model with few parameters and faster computations. Results on the case study demonstrate an average $45\%$ reduction in the computational need of the controller when compared against an MPC under the same environmental conditions. While PI-MPCS has certain advantages, it also has some limitation. As the reference trajectories $D$ may contain unfeasible samples, the \textit{discrete dynamics residual loss} and the \textit{Lyapunov stability loss} directly or indirectly reinforce the learning process of these unfeasible samples. As an result, all PI-MPCS trained with the physics-informed loss function(s) have lower safe landing rate (the ratio of trajectories that do not contain unfeasible samples) than the benchmark merely trained with the \textit{control loss}. In the future work, a filtering strategy will be developed, so that the unfeasible samples can be removed/modified without breaking the continuity of the trajectory. 


\bibliographystyle{unsrt}
\bibliography{root}

\end{document}